\documentclass[a4paper,11pt]{article}
\pdfoutput=1 % if your are submitting a pdflatex (i.e. if you have
\usepackage{jheppub} % for details on the use of the package, please
\usepackage{hyperref}
\usepackage{slashed}
\usepackage[T1]{fontenc} % if needed

\title{\boldmath Marginal deformations of a class of AdS$_3$ $\mathcal{N}=(0,4)$ holographic backgrounds}

\author[a]{Salomon Zacar\'ias}
\affiliation[a]{Department of Theoretical Physics and Astrophysics, Faculty of Science, Masaryk University\\611 37 Brno, Czech Republic}
\emailAdd{zacarias.salomon84@gmail.com}
\abstract{We discuss marginal deformations of warped AdS$_3\times$S$^2$ solutions preserving small $\mathcal{N}=(0,4)$ supersymmetry in massive IIA and eleven-dimensional supergravity and obtain
     a whole family of new solutions. We characterise these new backgrounds by studying
      some observables like the quantised charges, associated Hannany-Witten brane set-ups and the holographic central charge, the latter is shown to be invariant under the deformation. The study of the preservation of supersymmetry shows
       that the new backgrounds support an identity structure on the internal five-dimensional space, which is dynamical.}

\begin{document} 
\maketitle
\flushbottom

\section{Introduction and summary}
\label{intro}
 Supersymmetric solutions with AdS$_{\text{p}+1}$ factors in type II and eleven-dimensional supergravities play a prominent role in the context of the AdS/CFT correspondence since they provide a holographic description of p-dimensional superconformal field theories (SCFTs) at strong coupling \cite{Maldacena:1997re}. The discovery of this holographic duality has ever since triggered a number of efforts to construct and classify AdS vacua for any dimension allowed and preserving various amounts of (super)symmetries that have been used to study and characterise SCFTs.

 More recently, the case of AdS$_3$ backgrounds has gained a lot of attention. There are several motivations for this. 
  For instance, the near horizon geometries of five-dimensional extremal black holes have AdS$_3$ factors. Using SCFT$_2$ data it is then possible to understand microscopic features
   of black holes, like their entropy by computing the central charge of the SCFT$_2$ \cite{Strominger:1996sh}, among other aspects (see for instance  \cite{Maldacena:1997de,Minasian:1999qn,Castro:2008ne,Vafa:1997gr,Couzens:2019wls}). 
   On the other hand, two-dimensional SCFTs are special on their own since they can, in certain cases, be fully solvable due to the structure of the superconformal algebra. It is therefore interesting to explore deeply each side of this dual pair in order to shed some light on new phenomena via holography. For a sample of works regarding 
  AdS$_3$ supersymmetric backgrounds in ten and eleven-dimensional supergravity preserving different amounts of supersymmetry and their holographic applications see \cite{Martelli:2003ki,Figueras:2007cn,Donos:2008hd,Colgain:2010wb,Jeong:2014iva,Lozano:2015bra,Kelekci:2016uqv,Couzens:2017way,Couzens:2017nnr,Eberhardt:2017uup,Dibitetto:2018iar,Macpherson:2018mif,Legramandi:2019xqd,Lozano:2019emq,Lozano:2019jza,Lozano:2019zvg,Lozano:2019ywa,Couzens:2019mkh,Couzens:2019iog,Gauntlett:2006af,Filippas:2019ihy,Speziali:2019uzn,Lozano:2020bxo,Farakos:2020phe,Couzens:2020aat,Rigatos:2020igd,Faedo:2020nol,Dibitetto:2020bsh,Filippas:2020qku,Passias:2020ubv,Faedo:2020lyw,Eloy:2020uix,Deger:2019tem,Legramandi:2020txf}.
  %on the other hand, two-dimensional CFTs are interesting on their own since their structure make them fully solvable in certain cases.
  %two-dimensional CFTs  are special on their own due to their structure which in some cases make them fully solvable.
  
Moreover, on the geometrical side, attempts to constructing and classifying supersymmetric AdS$_3$ solutions have been mostly focused on the G-structure formalism \cite{Grana:2005sn} for which one extracts geometric constraints for the fields of the solutions according to the number
of (super)symmetries and geometrical structures, etc, we impose in the internal space. It has also been considered back-reacting D-brane arrangements which are known to produce 
AdS solutions in the near horizon limit \cite{Faedo:2020nol,Dibitetto:2020bsh}, among others.
However, these efforts have been non-exhaustive due to the many choices we have on the number of supersymmetries, and superconformal algebras, supported by the solutions constraining the internal space submanifolds.
Thus the approach has been focussed on searching and classifying all supergravity solutions preserving given amounts of supersymmetry, choices of internal structures, etc. 
This program has allowed to expand significatively our knowledge of new string backgrounds which may have very interesting applications in the context of holography.
In this vein, another possibility to explore the landscape of AdS$_3$ vacua is to consider AdS-preserving deformations of well-known supergravity solutions. Depending on the details of the deformation these solutions
may preserve supersymmetry whilst changing the structure of the internal space, and in some cases escape from presently known classifications of supergravity solutions.

  In this work we will use TsT transformations \cite{Lunin:2005jy} and the analog to eleven dimensions \cite{Gauntlett:2005jb} in order to generate a larger class of warped AdS$_3$ supersymmetric solutions. The seed backgrounds we will consider are a subclass of the solutions constructed in \cite{Lozano:2019emq} which are solutions of massive IIA supergravity of the warped form AdS$_3\times$S$^2\times$CY$_{2}$ foliated over an interval, the two-sphere realising geometrically the SU(2)$_{\text{R}}$ R-charge of the solution.
  They are given in terms of three linear functions, preserve small $\mathcal{N}=(0,4)$ supersymmetry and an SU(2) structure in the internal five-dimensional space, analogously, for these solutions, an SU(3) structure in the seven-dimensional space transverse to AdS$_3$. The above solutions appear in the near horizon limit of D$_2$-D$_4$-NS5-D$_6$-D$_8$ brane arrangements. D$_2$ and D$_6$ branes are colour branes and are suspended between the NS5 branes whilst D$_4$ and D$_8$ correspond to localised sources and provide flavour groups attached to the gauge nodes which leave the dual quiver CFT 
  anomaly free \cite{Lozano:2019jza,Lozano:2019zvg}.  For vanishing Romans mass, the uplift to eleven dimensions of the above solution gives rise to a class of AdS$_3\times \text{S}^3/\text{Z}_k\times$CY$_2$ foliated over an interval, which preserve the same amount of supersymmetry and internal structure group \cite{Lozano:2020bxo}. The brane configuration for this solution involves M2 branes and KK monopoles suspended between M5'  branes as well as 
  extra flavour M5 branes. 
  
Given the internal symmetries of the seed solutions above, we have two choices which produce inequivalent backgrounds after TsT transformations. Namely, if we consider or not the azimutal direction inside the S$^2$ for the process. 
In the latter case we are left with solutions for which supersymmetry is fully preserved. Of course more generic supersymmetric solutions can be generated by a sequence of TsT's not involving the U(1) inside the S$^2$,  but we will explore this more generic case in the future. 
%The backgrounds obtained in ten and eleven dimensions define a whole family of warped AdS$_3$ solutions, which can be used to generate the deformed-analog AdS$_2$ solutions studied in [] after double analytical continuation and reduction in the case of the eleven-dimensional solutions. 
For the ten-dimensional solutions, we study the brane configurations that we propose generate our solutions in the near-horizon limit.
Holographically, these new backgrounds are dual to marginal deformations of the  seed (undeformed) SCFT, both theories having the same central charge in the holographic limit. 
The latter can be understood since their degrees of freedom, in the aforementioned limit, are associated to the weighted volume of the internal spaces (to be defined below).
 The deformation changed the internal space enriching the geometric structure but left invariant its weighted volume. We prove this using the holographic calculation and left the specification of the SCFT for a forthcoming publication. 

 The content of this paper is organised as follows.  In Section \ref{solutionsetc} we start by briefly reviewing the seed solutions in \cite{Lozano:2019emq}. We then proceed to apply the TsT transformation in Section \ref{deformations10d} in order to obtain the new family of backgrounds in massive IIA. We study the quantised charges and present brane configurations which we argue give rise to our solutions in the near horizon limit. In Section \ref  {section11ddef} we study the eleven-dimensional analog of the TsT transformation for the solutions in Section \ref{solutionsetc} with vanishing romans mass uplifted to eleven dimensions.  One of the solutions obtained correspond to the uplift of the TsT-deformed IIA solution in the massless case.
  We then prove the invariance of the central charge under the deformation in Section \ref{sectionccharge} using the holographic computation. 
 Finally, In Section \ref{secsupersy} we study the preservation of supersymmetry for the solutions obtained in Sections \ref{deformations10d} and \ref{section11ddef} . This analysis suggest  the new supersymmetric solutions support a dynamical identity structure in the internal five-dimensional space. Some comments and final remarks are addressed in Section \ref{conclusions}. In Appendix \ref{supersymmetry}  we give our conventions for supersymmetry.

 % \section{Marginal deformations of a class of $\mathcal{N}=(4,0)$ holographic backgrounds}
  \section{The seed AdS$_3$ $\mathcal{N}=(0,4)$ holographic backgrounds}
  \label{solutionsetc}
 In this section we shall briefly review the AdS$_3$ solutions in massive IIA supergravity preserving small $\mathcal{N}=(0,4)$ supersymmetry obtained in \cite{Lozano:2019emq}. 
They will constitute our starting point from which  we will obtain the marginally deformed solutions via a transformation involving dualities. 

The solutions in \cite{Lozano:2019emq} are of the warped form AdS$_3\times$ S$^2\times$ M$_5$, supporting an SU(2) structure on M$_5$, equivalently, for these solutions,
an SU(3) structure in seven dimensions. Moreover, the five-dimensional space M$_5$ locally splits into a four-dimensional piece M$_4$ and an interval. There are two classes of solutions.
  In this work we will concentrate on a subclass of  {\it class} I solutions for which M$_4$ is (conformally) CY$_2$. From now on we will consider CY$_2=$T$^4$.
    The NS sector of the solution in  the string frame reads

  \begin{equation}
  \begin{split}\label{back1}
ds^2&= \frac{u}{\sqrt{h_4 h_8}}\bigg(ds^2_{\text{AdS}_3}+\frac{h_8 h_4 }{f_1}ds^2_{\text{S}^2}\bigg)+ \sqrt{\frac{h_4}{h_8}}ds^2_{T^4}+ \frac{\sqrt{h_4 h_8}}{u} d\rho^2, \\
&\qquad \qquad e^{\Phi}= \frac{2h_4^{1/4}}{h_8^{3/4}}\sqrt{\frac{u}{f_1}},~~~~~~ B_2= f_2\,d \text{vol}_{\text{S}^2}, 
\end{split}
\end{equation}
where the functions $u, h_4, h_8$ are functions of $\rho$ only. This is
supported with the following RR field strengths 
  \begin{equation}
  \begin{split}\label{back1fluxes}
F_0&=h_8',\;\;\;\;
F_2=-\frac{1}{2}\left(h_8-\frac{h'_8 u u'}{f_1}\right)d \text{vol}_{\text{S}^2},\quad F_8=\star_{10} F_2,\quad F_{10}=-\star_{10} F_0, \\ 
F_4&=- \bigg(\left(\frac{u u'}{2 \hat{h}_4}\right)'+2 h_8\bigg)  d\rho \wedge d\text{vol}_{\text{AdS}_3}
- h'_4\, d\text{vol}_{\text{T}^4}, \quad  F_6=-\star_{10} F_4, 
\end{split}
\end{equation}
where $'=\partial_{\rho}$ and
\begin{equation}
f_1=4 h_4 h_8+(u')^2,\quad f_2=\frac{1}{2}\left(-\rho+\frac{ u u'}{f_1} \right).
\end{equation}
The above background is a supersymmetric solution of massive type IIA supergravity provided 
\begin{equation}\label{conds}
h''_4(\rho)=0,\quad h''_{8}(\rho)=0,\quad u''(\rho)=0,
\end{equation}
the first two
away from localised sources. 
The $\rho$ coordinate parametrising the interval can be taken to be of finite range. This imposes additional constraints on the various functions of the solution. We require 
for $0\leq \rho\leq 2\pi (P+1)$ that \footnote{For other choices where some of these conditions are relaxed see \cite{Lozano:2019zvg,Filippas:2020qku}.}
\begin{equation}\label{bothcond}
h_8\vert_{\rho=0}=h_8\vert_{\rho=2\pi(P+1)}=h_4\vert_{\rho=0}=h_4\vert_{\rho=2\pi(P+1)}=0. 
\end{equation}
The metric functions obeying the above conditions are then explicitly given in Table \ref{metricfunctions}. 

 \begin{table}[h!]
\centering
 \begin{tabular}{c||c c c|ll|lc c c|c lc c c ll c c}
  & $0\leq \rho\leq 2\pi$ & $2\pi j\leq \rho\leq 2\pi(j+1)$ & $2\pi P\leq \rho\leq 2\pi(P+1)$  \\ [0.5ex] 
 \hline\hline %-----------------------------------------------------------------------------------------------
 $h_8$ & $ \frac{\nu_0 }{2\pi}\rho$ & $ \mu_j+\frac{\nu_{j}}{2\pi}(\rho-2\pi j)$ & $ \mu_P - \frac{\mu_P}{2\pi}(\rho-2P\pi )$\\ 
 \hline
 $h_4$ & $ \frac{\beta_0 }{2\pi}\rho$ & $ \alpha_j+\frac{\beta_{j}}{2\pi}(\rho-2\pi j)$ & $ \alpha_P - \frac{\alpha_P}{2\pi}(\rho-2 P\pi )$  \\ 
 \hline %-----------------------------------------------------------------------------------------------
  $u$ & &   $\frac{b_{0}}{2\pi}\rho$ & \\
  \hline
\end{tabular}
\caption{Piece-wise continuous functions satisfying the conditions in eq. \ref{bothcond}. The value of $u(\rho)$ is the same in all intervals, as required by supersymmetry.}
\label{metricfunctions}
\end{table}

The set of constants $(\alpha_j,\beta_j,\mu_j,\nu_j,b_0)$ for $j=0,\ldots P$  parametrising the piece-wise continuous functions above are subject to certain constraints imposing continuity of the NS sector along the $\rho$ intervals. 
The conditions are
\begin{equation}
\alpha_{k}=\sum_{j=0}^{k-1}\beta_j,\quad \mu_{k}=\sum_{j=0}^{k-1}\nu_j. 
\end{equation}
  The supergravity solution is trustable whenever these constants as well as the number P have large values.

 \section{The marginally deformed backgrounds}\label{deformations10d}
  In this section we will construct a family of solutions corresponding to deformations of the supergravity solutions in eqs (\ref{back1})-(\ref{back1fluxes}). 
  Such deformations are built upon a sequence of T dualities and a change of coordinates \cite{Lunin:2005jy}. The resulting backgrounds are considered to be 
  holographic duals of the marginally deformed SCFTs dual to the original (undeformed) backgrounds.
  
  In order to proceed, we first pick a two-torus in the geometry. For the solution in eq. (\ref{back1}) there are two options which will produce inequivalent solutions. They correspond to U(1)$_{\varphi}\times$ U(1)$_{x_{i}}$
  and U(1)$_{x_{i}}\times$ U(1)$_{x_{j}}$ invariant sub-sectors, where $\varphi$ is the azimuthal angle inside the S$^2$ and  $x_{i}$ the coordinates on T$^4$. The deformation is achieved by performing a T-duality in one of the coordinates, a shift with parameter $\lambda$
  in the second and T duality back in the first. The solutions obtained will describe a family of solutions in terms of the functions $u,h_4, h_8$ and the parameter $\lambda$. 
  
 In the first case T$^{2}: (\varphi,x_1)$, following the T duality rules in \cite{Kachru:2002sk} , the above procedure generates the following background
  \begin{align}\label{back2}
  ds^2&=  \frac{u}{\sqrt{h_4 h_8}}\bigg(ds^2_{\text{AdS}_3}+\frac{h_8 h_4 }{f_1}d\theta^2\bigg)+\sqrt{\frac{h_4}{h_8}}(dx_2^2+dx_3^2+dx_4^2)+\frac{\sqrt{h_4 h_8}}{u} d\rho^2\nonumber\\
  &+\frac{1}{f_1+\lambda^2\sin^2\theta h_4 u}\left(u\sqrt{h_4 h_8}\sin^2\theta d\varphi^2+\sqrt{\frac{h_4}{h_8}}f_1(dx_1-\lambda f_2\sin\theta d\theta)^2\right),\nonumber\\
 &~~~~~~~~~~~~~~~~~  e^{\Phi}=\frac{2 h_4^{1/4}}{h_8^{3/4}}\sqrt{\frac{u}{f_1+\lambda^2h_4 u \sin^2\theta}},\\
  &\qquad B_2=\frac{\lambda h_4 u \sin^2\theta}{f_1+\lambda^2 h_4 u \sin^2\theta}(dx_1-\lambda f_2\sin\theta d\theta)\wedge d\varphi+f_2d\text{vol}_{ \text{S}^2},\nonumber\\
  F_0=&h_8',\qquad F_2=\frac{\gamma h_4 u\sin^2\theta h_8'}{f_1+\lambda^2h_4 u\sin^2\theta}(dx_1-\lambda f_2\sin\theta d\theta)\wedge d\varphi-\frac{1}{2}\left(h_8-\frac{h'_8 u u'}{f_1}\right)d\text{vol}_{ \text{S}^2}\nonumber\\
  F_4=& -\bigg(\left(\frac{u u'}{2 \hat{h}_4}\right)'+2 h_8\bigg)  d\rho \wedge d\text{vol}_{\text{AdS}_3}
- h'_4d\text{vol}_{\text{T}^4}+\frac{1}{2}\gamma (h_4-\rho h_4')\sin\theta d\theta\wedge dx_2\wedge dx_3\wedge dx_4\nonumber, 
\end{align}
where the higher fluxes are obtained via the lower ones as indicated in eq (\ref{back1fluxes}). The background in eq. (\ref{back2})
 is a solution of massive IIA supergravity if conditions in eq. (\ref{conds}) are imposed. We notice the original solution is recovered after turning off the
deformation parameter, as expected.

For the second case T$^2: (x_3,x_4)$, the procedure outlined above produces the following background 
 \begin{equation}
  \begin{split}\label{back3}
&  ds^2= \frac{u}{\sqrt{h_4 h_8}}\bigg(ds^2_{\text{AdS}_3}+\frac{h_8 h_4 }{f_1}ds^2_{S^2}\bigg)+ \frac{\sqrt{h_4 h_8}}{u} d\rho^2 
 +\sqrt{\frac{h_4}{h_8}}dx_{1,2}^2+ \sqrt{\frac{h_4}{h_8}}\frac{1}{1+\lambda^2\frac{h_4}{h_8}}dx_{3,4}^2,\\ 
& e^{\Phi}= \frac{2h_4^{1/4}}{h_8^{3/4}}\sqrt{\frac{u}{(1+\lambda^2\frac{h_4}{h_8})f_1}},~~~~~~ B_2= f_2d\text{vol}_{ \text{S}^2}-\lambda \frac{h_4}{h_8}\frac{dx_3\wedge dx_4}{1+\lambda^2\frac{h_4}{h_8}},\\
&F_0=h_8',\;\;\;\;
F_2=-\frac{1}{2}\left(h_8-\frac{h'_8 u u'}{f_1}\right)d\text{vol}_{ \text{S}^2}-\lambda h_4' dx_1\wedge dx_2-\lambda \frac{h_4 h'_{8}}{h_8(1-\lambda^2 \frac{h_4}{h_8})}dx_3\wedge dx_4,\\
&F_4= -\bigg(\left(\frac{u u'}{2 h_4}\right)'+2 h_8\bigg)  d\rho \wedge d \text{vol}_{\text{AdS}_3}
-\frac{h'_4}{1+\lambda^2\frac{h_4}{h_8}}d \text{vol}_{\text{T}^4}+\\
&\lambda \left( \frac{ h_4}{2 h_8(1+\lambda^2\frac{h_4}{h_8})}\left(h_8-\frac{u u' h_8'}{f_1}\right)dx_3\wedge dx_4+\frac{1}{2}\left(h_4-\frac{u u' h_4'}{f_1}\right)dx_1\wedge dx_2\right) \wedge d\text{vol}_{\text{S}^2},
\end{split}
\end{equation}
which is a solution of massive IIA supergravity if conditions in eq. (\ref{conds}) are imposed. We notice since the S$^2$
is a spectator subspace for this deformation, we expect $\mathcal{N}=(0,4)$ supersymmetry will be fully preserved as we will
explicitly show in Section \ref{secsupersy}. 

\subsection{Quantised charges and brane set-ups}\label{charges}
In this section we will study the Page charges of the deformed backgrounds. Throughout, we shall use the following definitions \footnote{we will be using units such that $g_s=\alpha'=1$ } $Q_{D_p}=\frac{1}{(2\pi)^{7-p}}\int_{_{\Sigma_{8-p}}} \hat{f}_{8-p}$, where $\hat{f}=e^{-B\wedge } f$, here $f$ denotes the magnetic (internal) part of the RR polyform $F$.

We start with the solution in eq. (\ref{back2}) and consider the following non-trivial cycles of the geometry
% \footnote{The solutions we are getting are considering LGT, so  we have to decide where to add the discussion of LGT. Notice the $\gamma$ dependent term does not contribute to new flux of NS5, so LGT are then only discussed as in the usual case on the $S^2$. }
\begin{equation}
\begin{split}
\Sigma_{_{2}}=&S^2,\quad  \Sigma_{_{4}}=T^4,\quad \Sigma_{_{3}}=(\rho,S^2),\quad \Sigma_{_{6}}=(S^2, T^4),
 \quad \Sigma_{_{4}}'=(x_2,x_3,x_4,\theta),\quad \Sigma_{_{3}}'=(x_1,\rho,\varphi).
\end{split}
\end{equation}
The Page charges read 
\begin{equation}\label{page1}
\begin{split}
Q_{_{\text{NS}5}}=&\frac{1}{(2\pi)^2}\int_{ \Sigma_3}H_3,\qquad Q_{_{\text{NS}5'}}=\frac{1}{(2\pi)^2}\int_{\Sigma_3'}H_3,\quad  Q_{D_8}=2\pi h_8', \quad Q_{D_6}=\frac{1}{2\pi}\int_{ \Sigma_2}\hat{f}_2\\
Q_{D_4}=&\frac{1}{(2\pi)^3}\int_{ \Sigma_4}\hat{f}_4,\qquad Q_{D'_4}=\frac{1}{(2\pi)^3}\int_{ \Sigma_4'}\hat{f}_4,\quad Q_{D_2}=\frac{1}{(2\pi)^5}\int_{\Sigma_6}\hat{f}_6.
\end{split}
\end{equation} 
       If we allow large gauge transformations $B_2\rightarrow B_2+\pi k\, d\text{vol}_{\text{S}^2}$, the Page fluxes are those in eq. (\ref{back2}) except for
           \begin{equation}
       \begin{split}
       \hat{f_2}=&-\frac{1}{2}(h_8-h_8'(\rho-2\pi k)) d\text{vol}_{\text{S}^2},\\
         \hat{f_6}=&\frac{1}{2}(h_4-h_4'(\rho-2\pi k)) d\text{vol}_{\text{S}^2} \wedge d\text{vol}_{\text{T}^4}. 
       \end{split}
       \end{equation}
        The charges in eq. (\ref{page1}) computed in $2\pi k\leq \rho \leq2\pi (k+1)$ are explicitly, 
        \begin{equation}
        \begin{split}
        Q_{D_8}=&\nu_k,\quad Q_{D_6}=\mu_k,\quad
                Q_{D_4}=\beta_k,\quad 
                Q_{D_2}=\alpha_k,\\
  &Q_{D_4'}=\lambda(k\beta_k-\alpha_k)\quad Q_{_{\text{NS}5}}=1,\quad Q_{_{\text{NS}5'}}=0.
        \end{split}
        \end{equation}
        We notice that for finite $\rho\in[0,2\pi(P+1)]$ 
        we have $P+1$ parallel NS5 branes.
        From the above expressions we see in particular that no extra NS5' branes were generated by the deformation. In addition, 
        the above charges are well-defined as long as the set of constants $\alpha_k,\beta_k,\mu_k,\nu_k$ as well as the combination $\lambda(k\beta_k-\alpha_k), \in\mathbb{Z}$.

   As we pointed out before, the first two conditions in eq. (\ref{conds}) must be satisfied by the solutions everywhere except at points were we have localised sources. 
   At those points, we have a change in gradient of the piece-wise linear functions proportional to $h''_{4,8}$ pointing the possible existence of  a source for D$_p$ branes via
      the modified Bianchi identities $d\hat{f}=j_{s}$. From Table \ref{metricfunctions} we obtain 
      \begin{equation}\label{derivatives}
      h''_4=\sum_{k=1}^{P}\left(\frac{\beta_{k-1}-\beta_k}{2\pi}\right)\delta(\rho-2\pi k ), \qquad        h''_8=\sum_{k=1}^{P}\left(\frac{\nu_{k-1}-\nu_k}{2\pi}\right)\delta(\rho-2\pi k ).
      \end{equation}
      
Using this information as well as the Page fluxes of the solution we compute
        \begin{align}
        d\hat{f}_0=&2\pi h''_{8}d\rho,\\
        d\hat{f_2}=&\frac{1}{2}(\rho-2\pi k)h''_8d\rho\wedge d\text{vol}_{\text{S}^2}=0,\\
        d\hat{f_4}=&h''_{4}d\rho\wedge\left( d\text{vol}_{\text{T}^4}+\lambda\frac{\rho}{2}\sin\theta d \theta\wedge dx_2\wedge dx_3 \wedge dx_4\right)\\
        d\hat{f}_6=&\frac{h_4''}{2}(\rho-2\pi k)d\rho\wedge d\text{vol}_{\text{S}^2} \wedge d\text{vol}_{\text{T}^4}=0
        \end{align}
        where we have used $x\delta(x)=0$. From this we conclude that D$_4$, D$'_4$ as well as D$_8$, having non-zero sources, correspond to flavour branes
      whilst D$_6$ and D$_2$ are colour ones. 
       Thus in addition to the D-branes of the seed solution, the deformation has induced (semi-localised) flavour $Q_{\text{D}'_4}$ branes. The brane configuration, before the near horizon limit is taken, we argue is associated to the solution above 
       is shown in Table \ref{tableblabla1}. 
       %This can be thought of as deforming first the space realising the internal isometries of the seed solution where the seed brane arrangement will then be placed. 
       %this can be thought of as the seed brane configuration placed in the deformed space realising the internal isometries of the solution
       %can be thought of as placing the branes in the deformed space realising the isometries of the solution 
      % where the seed brane configuration was placed, then placing the branes on that deformed space.  

        %The general idea is that this Hannany-Witten setup, in the near horizon limit, generates the solution we obtained for large values of the constants. This is generically true.
 %However, in our case the branes can be thought of as placed on a deformed flat space which I do not know if the analysis/interpretation can go through
              \begin{table}[h!]
\centering
 \begin{tabular}{c|c c|lcc|lccc|c| cl cllcll}
  & $t$ & $x$ & $r$ & $\theta$ & $\phi$ & $x_1$ & $x_2$ & $x_3$ & $x_4$ & $\rho$ & N$_{\text{D}_{p}}$ \\ [0.5ex] 
 \hline\hline %-----------------------------------------------------------------------------------------------
 D$_8$ & $\bullet$ & $\bullet$ & $\bullet$ & $\bullet$ & $\bullet$ & $\bullet$ & $\bullet$ & $\bullet$ & $\bullet$ & $\cdot$ & $\nu_{k-1}-\nu_{k}$\\ 
 \hline
 D$_6$ & $\bullet$ & $\bullet$ & $\cdot$ & $\cdot$ & $\cdot$ & $\bullet$ & $\bullet$ & $\bullet$ & $\bullet$ & $\bullet$ & $\mu_k$\\ 
 \hline\hline %-----------------------------------------------------------------------------------------------
  D$'_4$ & $\bullet$ & $\bullet$ & $\bullet$ & $\cdot$ & $\bullet$ & $\bullet$ & $\cdot$ & $\cdot$ & $\cdot$ & $\cdot$ & $\lambda k \text{N}_{D_4}$ \\
 \hline\hline %-----------------------------------------------------------------------------------------------
  D$_4$ & $\bullet$ & $\bullet$ & $\bullet$ & $\bullet$ & $\bullet$ & $\cdot$ & $\cdot$ & $\cdot$ & $\cdot$ & $\cdot$ & $\beta_{k-1}-\beta_{k}$ \\ 
 \hline
  D$_2$ & $\bullet$ & $\bullet$ & $\cdot$ & $\cdot$ & $\cdot$ & $\cdot$ & $\cdot$ & $\cdot$ & $\cdot$ & $\bullet$ &  $\alpha_k$\\
  \hline
  NS5 & $\bullet$ & $\bullet$ & $\cdot$ & $\cdot$ & $\cdot$ & $\bullet$ & $\bullet$ & $\bullet$ & $\bullet$ & $\cdot$ & 
\end{tabular}
\caption{Brane configuration which in the near horizon limit gives the solution in eq. (\ref{back2}). We show the world-volume directions the branes are suspended as well as their number in the k-th interval.}
\label{tableblabla1}
\end{table}

        For the second solution, we consider the following cycles
        \begin{equation}
        \Sigma_{_{2}}'=(x_1,x_2),\quad \Sigma_{_{4}}=T^4,\quad \Sigma'_{_{4}}=(\text{S}^2,x_1,x_2),\quad \Sigma_{_{6}}=(\text{S}^2,\text{T}^4),\quad \Sigma_{_{3}}'=(\rho,x_3,x_4),
        \end{equation}
        
    and non-trivial Page forms
         \begin{equation}
        \begin{split}
        \hat{f_2}=&-\frac{1}{2}(h_8-h_8'(\rho-2\pi k))d\text{vol}_{\text{S}^2}+\lambda\, h_4' dx_1\wedge dx_2,\\
        \hat{f}_4=&h'_4d \text{vol}_{\text{T}^4}
        -\frac{\lambda}{2}(h_4-h_4'(\rho-2\pi k))d\text{vol}_{\text{S}^2}\wedge dx_1\wedge dx_2,\\
        \hat{f_6}=&-\frac{1}{2}(h_4-h_4'(\rho-2\pi k))d\text{vol}_{\text{S}^2} \wedge d\text{vol}_{\text{T}^4}.
        \end{split}
        \end{equation}
       An analysis as detailed above shows that in addition to the D-branes of the seed solution,  the generated Page charges after the transformation ($\lambda$-dependent) are given by 
        \begin{equation}
        Q_{\text{D}'_6}=\lambda \beta_k,\quad Q_{\text{D}'_4}=\lambda \alpha_k.
        \end{equation}
        This implies the quantisation conditions $\lambda\beta_k\in \mathbb{Z}, \lambda\alpha_k\in \mathbb{Z}$, which requieres rational $\lambda$. 
        In order to determine if the above charges correspond to colour or flavour branes, we compute 
       \begin{align}
        d\hat{f_2}=&\lambda h''_4 d\rho\wedge dx_{1}\wedge dx_{2}+\frac{1}{2}(\rho-2k\pi)h_8''d\rho \wedge d\text{vol}_{\text{S}^2}\\
        d\hat{f_4}=&h_4'' d\rho\wedge d\text{vol}_{\text{T}^4}-\frac{\lambda}{2}(\rho-2k\pi)h_4''d\rho\wedge d\text{vol}_{\text{S}^2}\wedge dx_1\wedge dx_2.
        \end{align}
        Using then (\ref{derivatives}) we find that the effect of the deformation was to add $Q_{D'_4}$ colour and $Q_{D'_6}$ flavour branes respectively.  
        % We see that D$_4\rightarrow$ D$_6$, D$_2\rightarrow$ D$_4$ after the T dualities (the same may happen with the higher order branes). 
        Therefore  the original D$_4$-NS5-D$_2$ and D$_8$-NS5-D$_6$ brane arrangements are modified by the addition of D$'_4$ branes extended along
        $(t,x,x_3,x_4,\rho)$ as well as semi-localised D$'_6$ branes in  $(\text{AdS}_3,\text{S}^2,x_3,x_4)$  wrapped on T$^{2}:(x_3,x_4)$. 
       The brane set-up corresponding to this configuration is summarised in Table \ref{tableblabla}.
 % We notice that the $\gamma$-dependent charge associated to the NS5 branes $Q_{\textrm{NS}}=\frac{1}{(2\pi)^2}\int_{_{\Sigma_5'}} H_3$ is ill defined due to the vanishing conditions of $h_4, h_8$ at the end of the $\rho$ interval.

 \begin{table}[h!]
\centering
 \begin{tabular}{c|c c|lcc|lccc|c| cl cllcll}
  & $t$ & $x$ & $r$ & $\theta$ & $\phi$ & $x_1$ & $x_2$ & $x_3$ & $x_4$ & $\rho$ & $N_{D_{p}}$ \\ [0.5ex] 
 \hline\hline %-----------------------------------------------------------------------------------------------
 D$_8$ & $\bullet$ & $\bullet$ & $\bullet$ & $\bullet$ & $\bullet$ & $\bullet$ & $\bullet$ & $\bullet$ & $\bullet$ & $\cdot$ & $\nu_{k-1}-\nu_{k}$\\ 
 \hline
 D$_6$ & $\bullet$ & $\bullet$ & $\cdot$ & $\cdot$ & $\cdot$ & $\bullet$ & $\bullet$ & $\bullet$ & $\bullet$ & $\bullet$ & $\mu_k$\\ 
 \hline\hline %-----------------------------------------------------------------------------------------------
  D$'_6$ & $\bullet$ & $\bullet$ & $\bullet$ & $\bullet$ & $\bullet$ & $\cdot$ & $\cdot$ & $\bullet$ & $\bullet$ & $\cdot$  & $\lambda\, N_{_{D_4}}$\\
  \hline
  D$'_4$ & $\bullet$ & $\bullet$ & $\cdot$ & $\cdot$ & $\cdot$ & $\cdot$ & $\cdot$ & $\bullet$ & $\bullet$ & $\bullet$ & $\lambda\,  N_{_{D_2}}$ \\
 \hline\hline %-----------------------------------------------------------------------------------------------
  D$_4$ & $\bullet$ & $\bullet$ & $\bullet$ & $\bullet$ & $\bullet$ & $\cdot$ & $\cdot$ & $\cdot$ & $\cdot$ & $\cdot$ & $\beta_{k-1}-\beta_{k}$ \\ 
 \hline
  D$_2$ & $\bullet$ & $\bullet$ & $\cdot$ & $\cdot$ & $\cdot$ & $\cdot$ & $\cdot$ & $\cdot$ & $\cdot$ & $\bullet$ &  $\alpha_k$\\
  \hline
  NS5 & $\bullet$ & $\bullet$ & $\cdot$ & $\cdot$ & $\cdot$ & $\bullet$ & $\bullet$ & $\bullet$ & $\bullet$ & $\cdot$ & 
\end{tabular}
\caption{Brane configuration which in the near horizon limit gives the solution in eq. (\ref{back3}). We also show the world-volume directions the branes are suspended as well as their number in the k-th interval. We see the D$'_6$ and D$'_4$ branes are wrapped on T$^2$ : $(x_3,x_4)$.}
\label{tableblabla}
\end{table}

\section{The deformation in eleven dimensions}\label{section11ddef}
    In this section we will study a generalisation to eleven dimensional supergravity of the TsT transformation studied in the previous section. The seed solutions 
    will be the uplift of the background in eq. (\ref{back1})-(\ref{back1fluxes}) for vanishing Romans mass.  The backgrounds obtained will correspond to a family of 
    supersymmetric solutions which are out of a subclass of the classification for AdS$_3$ eleven dimensional solutions studied in \cite{Lozano:2020bxo}. 
    
    In order to proceed, we consider a vanishing Romans mass in the solution of  eq. (\ref{back1})-(\ref{back1fluxes}) which lead us to consider $h_8=k$. 
    The uplift of this solution to eleven dimensions was first constructed in \cite{Lozano:2020bxo}. 
    For latter use we will present some details here. We determine the three and one-form potentials to be
        \begin{equation}
        A_3=\left(\frac{u u'}{2h_4}+2k \rho\right)\,d\text{vol}_{\text{AdS}_3}+h_4' x_1\, d\text{vol}_{\text{T}^3}, \quad A_1=\frac{k}{2}\cos\theta d\varphi.
        \end{equation}
        We notice the 3-form potential above is not globally well-defined. This would be the case if $h_4'$ were a continuous function.
     Using the usual KK anzats eq. (\ref{anpp}), the eleven-dimensional solution raeds
        \begin{equation}
        \begin{split}\label{sol11d}
        ds^{2}_{11}=&\Upsilon\left(\frac{u}{\sqrt{h_4 k}}ds^{2}_{\text{AdS}_3}+\sqrt{\frac{h_4}{k}}ds^2_{\text{T}^4}+\frac{\sqrt{h_4 k}}{u}d\rho^2+\frac{k^2}{\Upsilon^3}ds^2_{\text{S}^3/\text{Z}_{k}}\right),\\
		         G_4=&dC_3=d\left(\left(\frac{u u'}{2 h_4}+2\rho k\right)d\text{vol}_{\text{AdS}_3}+2 k \left(-\rho+2\pi k+\frac{u u'}{f_1}\right)d\text{vol}_{\text{S}^3/\text{Z}_{k}}\right)+h_4'd\text{vol}_{\text{T}^4},
\end{split}
        \end{equation}
        
        where 
        \begin{equation}
        \begin{split}\label{upup}
        ds^2_{\text{S}^3/\text{Z}_{k}}=\frac{1}{4}\left(ds^2_{S^2}+\left(\frac{2}{k} dx_{_{11}}+\cos\theta d\varphi\right)^2\right),\quad 
        \Upsilon=\frac{f_1^{1/3} \sqrt{k}}{(4 u\sqrt{h_4})^{1/3}}.
        \end{split}
        \end{equation}
        This solution preserves small $\mathcal{N}=(0,4)$ supersymmetry and supports an SU(2) structure. We will now generalise this class of solutions by performing 
        an SL(3,R) transformation of coordinates. 
        
          % where T$^3:(x_2.x_3.x_4)$ is the three torus on which an SL(3,R) transformation will be applied to generate the new background.

         % The brane configuration for this solution consists of an arrangement of M5', M5  and M2 as well as KK monopoles (see Table (\ref{tableblabla2})).  The first two appear from the magnetic components of the four-form flux in eq. (\ref{sol11d})
        %whilst the M2s appear as the flux through $(S^{3}/Z_k, T^4)$ of the $\hat{G}_7$ page form, $\hat{G_7}=-\star F_4-F_4\wedge C_3$. The orbifolding parameter, $k$, is identified with the KK monopoles (wrapped on $T^4$). This $M5'\vert MKK\vert M2\vert M5$ configuration defines $M$ strings probing A type singularities with extra $M5$ branes, which correspond to flavour membranes. The explicit solution involving the above brane intersection, the near horizon of which gives the background in eq. (\ref{sol11d}), was constructed in \cite{Faedo:2020nol}. 
        
  For a solution which is SL(3,R) invariant we use the anzats  
        \begin{equation}
\begin{split}\label{deformed}
&\quad ds_{_{11}}^2=\Delta^{-1/6}g_{\mu\nu}dx^{\mu}dx^{\nu}+\Delta^{1/3}M_{ab}\mathcal{D}\phi^{a}\mathcal{D}\phi^{b},\\
C_3=&C_{(0)}\mathcal{D}\phi^{1}\wedge\mathcal{D}\phi^{2}\wedge\mathcal{D}\phi^{3}+\frac{1}{2}C_{(1)ab}\wedge\mathcal{D}\phi^{a}\wedge\mathcal{D}\phi^{b}+C_{(2)a}\wedge\mathcal{D}\phi^{a}+C_{(3)},\\
&\qquad \qquad \qquad  \mathcal{D}\phi^{a}=d\phi^{a}+A_{\mu}^{a}dx^{\mu},
\end{split}
\end{equation}
where the $a,b$ indices correspond to the three-torus directions, $g_{\mu\nu} $ is the transverse eight dimensional metric and detM=1.
We have two possible choices for which we can apply the transformation. Namely T$^3$:$(x_2,x_3,x_4)$ and T$^3$:$(x_3,x_4,x_{11})$.

In the first case, the background in eq. (\ref{sol11d}) can be bring into the form of eq. (\ref{deformed}) provided we identify 
\begin{align}\label{deformed2}
&\qquad A_{\mu}^{a}=0,\quad 
M_{ab}=\Delta^{-1/3}\left(\frac{h_4 f_1}{4 u}\right)^{1/3}\delta_{ab},
\quad \Delta=\frac{h_4 f_1}{4 u}\nonumber\\
 C_{(0)}=&-h_4' x_1,\quad C_{(3)}=-\left(\frac{u u'}{2h_4}+2 k \rho\right)d\text{vol}_{\text{AdS}_3}+B_2\wedge dx_{_{11}},\nonumber\\ &\qquad \qquad \qquad C_{(1)ab}=C_{(2)a}=0, \\
\Delta^{-1/6}g_{\mu\nu}dx^{\mu}dx^{\nu}=&\sqrt{k} \left(\frac{ f_1}{4 \sqrt{h_4} u}\right)^{1/3}\left(\frac{u}{\sqrt{h_4 k}}\bigg(ds^2_{\text{AdS}_3}+\frac{ h_4 k }{f_1}ds^2_{\text{S}^2}\bigg)+ \sqrt{\frac{h_4}{k}}dx_{1}^2+ \frac{\sqrt{h_4 k}}{u}
d\rho^2\right)\nonumber\\+&\frac{k}{4}\left(\frac{16 u^2 h_4}{f_1^2}\right)^{1/3}(\frac{2}{k}dx_{_{11}}+\cos\theta d\varphi)^2\nonumber, 
\end{align}
We then use the transformation rules spelled out in \cite{Gauntlett:2005jb} to obtain the new background parametrised by $\lambda$. The transformation for the one-form $A^{a}$, using (\ref{deformed2}),  gives  $\tilde{A}^{a}=A^{a}+\frac{1}{2}\lambda \epsilon^{abc}C_{(1)bc}=0$ and therefore $\tilde{\mathcal{D}}\phi^{a}=\mathcal{D}\phi^{a}=d\phi^{a}$.
On the other hand, the non-trivial transformation associated to $\tau=-C_{(0)}+i\Delta^{1/2}$ reads $\tilde{\tau}=\tau/(1+\lambda\tau)$, from which we obtain 
 
\begin{equation}
\tilde{\Delta}=G^2\Delta, \quad \tilde{C}_{(0)}=G\left(C_{(0)}-\lambda(C_{(0)}^2+\Delta)\right), \quad G=(1+2\lambda C_{(0)}-\lambda^2(C_{(0)}^2+\Delta))^{-1}.
\end{equation}
The deformed background then reads
\begin{equation}\label{defin11d}
\begin{split}
& ds^{2}_{11}=G^{-1/3}\left[\Upsilon\left(\frac{u}{\sqrt{h_4 k}}ds^{2}_{\text{AdS}_3}+\sqrt{\frac{h_4}{k}}dx_1^2+\frac{\sqrt{h_4 k}}{u}d\rho^2+\frac{k^2}{\Upsilon^3}ds^2_{\text{S}^3/\text{Z}_{k}}\right)+G\Upsilon\sqrt{\frac{h_4}{k}}ds^2_{\text{T}^3}\right],\\
&\qquad \qquad \qquad \qquad \qquad \qquad \qquad\qquad   G_{4}=d C_{3}, 
\end{split}
\end{equation}
where 
%=(1-\gamma C_{(0)})\, d\left((\frac{u u'}{2 h_4}+2 k \rho)d\text{vol}_{\text{AdS}_3}+4 k f_2 d\text{vol}_{\text{S}^3/\text{Z}_{k}}\right)+d(\tilde{C}_{0}d\text{vol}_{\text{T}^3}) -\gamma\sqrt{\Delta}\star_{8} F_{(4)},\\
\begin{equation}
\begin{split}
C_3=&-\left(\frac{u u'}{2 h_4}+2 k \rho+\lambda\frac{x_1}{2h_4}\left(h_4 f_1+u u' h'_4\right)\right)d\text{vol}_{\text{AdS}_3}+\tilde{C}_0\, d\text{vol}_{\text{T}^3}\\
&+4 k\left(f_2-\lambda\frac{x_1}{2}\left(h_4-\frac{u u' h'_4}{f_1}\right)\right) d\text{vol}_{\text{S}^3/\text{Z}_{k}}.\\
&
\end{split}
\end{equation}
This background is a solution of 11d supergravity when conditions in eq. (\ref{conds}) are imposed,
and reduces to the undeformed solution for $\lambda=0$, as expected.

In the second case T$^3:(x_3,x_4,x_{11})$, the solution obtained following the procedure spelled out above corresponds to the uplift to eleven dimensions of the solution in eq. (\ref{back3}).
The eleven-dimensional background reads 
        \begin{equation}
        \begin{split}\label{sol11ddeformed}
        ds^{2}_{11}=&\Upsilon(1+\lambda^2\frac{h_4}{h_8})^{1/3}\left(\frac{u}{\sqrt{h_4 c_8}}ds^{2}_{\text{AdS}_3}+\frac{h_8 h_4}{f_1}ds^2_{\text{S}^2}+\frac{\sqrt{h_4 h_8}}{u}d\rho^2+\sqrt{\frac{h_4}{h_8}}(dx_1^2+dx_2^2)+\right.\\
        +&\left.\sqrt{\frac{h_4}{h_8}}\frac{1}{1+\lambda^2\frac{h_4}{h_8}}(dx_3^2+dx_4^2)+\frac{c_8^2}{4\Upsilon^3}\left(ds^2_{\text{S}^2}+\frac{Dy^2}{1+\lambda^2\frac{h_4}{h_8}}\right)\right),\\
		         G_4=&-\left(\partial_{\rho}\left(\frac{u u'}{2 h_4}\right)+2 c_8 \right)d\rho\wedge d\text{vol}_{\text{AdS}_3}-\frac{h_4'}{1+\lambda^2\frac{h_4}{c_8}}d\text{vol}_{\text{T}^4}-\frac{\lambda}{2}\frac{h'_4}{\left(1+\lambda^2\frac{h_4}{c_8}\right)^2} d\rho\wedge d x_3\wedge dx_4 \wedge Dy\\
		         +&\frac{\lambda}{2}\left(\frac{h_4}{1+\lambda^2\frac{h_4}{c_8}}dx_{3}\wedge dx_{4}+\left(h_4-\frac{u u' h'_4}{f_1}\right) dx_1\wedge dx_2\right) \wedge d\text{vol}_{\text{S}^2}+\frac{c_8}{2}\partial_{\rho}f_2 d\rho\wedge d\text{vol}_{\text{S}^2}\wedge Dy,
\end{split}
        \end{equation}
where 
\begin{equation}
Dy=\frac{2}{c_8} dy+\cos\theta d\varphi-\frac{2}{c_8}\lambda x_1 h'_4 dx_2, 
\end{equation} and $\Upsilon$ was defined in eq. (\ref{upup}). This background is a solution of 11d supergravity when conditions in eq. (\ref{conds}) are imposed. 
In Section \ref{secsupersy} we will show that the solutions presented in this section preserve $\mathcal{N}=(0,4)$ supersymmetry supporting an identity structure.

Before to close this section, it is worth noticing that the solutions in eqs. (\ref{defin11d}) and (\ref{sol11ddeformed}) can be used as seed solutions in order to generate other families of supersymmetric solutions.
For instance, after appropriate analytical continuations we can generate solutions with $\text{AdS}_3/\text{Z}_k\times \text{S}^3$ factors which further reduction to IIA along the Hopf-fibre direction of AdS$_3$ 
will generate new AdS$_2\times\text{S}^3$ solutions in IIA supergravity, which can be further extended to massive IIA, 
%supporting an identity structure in the internal five-dimensional space transverse to AdS$_2\times$S$^3$, 
 generalising those studied in \cite{Lozano:2020bxo}, etc.

 \section{Holographic central charge}\label{sectionccharge}
The main goal of this section will be to compute the central charge characterising the new family of solutions. For the seed solutions this was done in \cite{Lozano:2019zvg,Lozano:2020bxo} and using the analysis of the spin-2 spectrum in \cite{Speziali:2019uzn}.
A generic result involving the deformations discussed above is that they leave the internal space volume transverse to AdS$_3$ -weighted by the dilaton- invariant. We then anticipate 
the central charges will be the same before and after the deformation. 

In order to see this explicitly, we consider the metric of the solutions written in the following way
\begin{equation}\label{mcentralc}
ds^2=a(r,\vec{y}î)\left(dx_{1,d}^2+b(r)dr^2\right)+g_{ij}(r,\vec{y})dy^{i}dy^{j}, 
\end{equation}
where $x_{1,d}$ parametrises $\mathcal{M}^{1,d}$ Minkowski space and $g_{ij}$ the metric of the internal space. 
The holographic central charge is then given by the following expression \cite{Macpherson:2014eza}
\begin{equation}
c_{_{\text{hol}}}=\frac{d^d}{G_{N}}\frac{b(r)^{d/2}H^{\frac{2d+1}{2}}}{H'^{d}},
\end{equation}
where 
\begin{equation}
H=\left(\int d\vec{y}\sqrt{e^{-4\Phi}\text{det}(g_{ij})a^{d}}\right). 
\end{equation}
For the ten dimensional solution, since the deformations acted on the internal space of the solutions, we clearly see the quantities $a(r,\vec{y}î),b(r)$ in eq. (\ref{mcentralc}) are spectator under the deformations. 
In addition, we find that 
\begin{equation}
e^{\tilde{\Phi}}=\frac{e^{\Phi}}{\sqrt{1+\lambda^2\text{det}(g_{_{T^2}})}}, \quad \tilde{g}_{_{T^2}}=\frac{g_{_{T^2}}}{1+\lambda^2\text{det}(g_{_{T^2}})}
\end{equation}
where tilde denotes fields after the deformation. It is then easy to see that $e^{-4\tilde{\Phi}}\text{det}(\tilde{g}_{ij})\tilde{a}^d=e^{-4\Phi}\text{det}(g_{ij})a^d$, giving $\tilde{c}_{_{\text{hol}}}=c_{_{\text{hol}}}$
as anticipated. This result goes through for the eleven-dimensional solutions after considering the relation between the ten and eleven-dimensional quantities in the KK anzats  (\ref{anpp}) and 
$H=\left(\int d\vec{\hat{y}}\sqrt{\text{det}(\hat{g}_{ij})\hat{a}^{d}}\right)$, where quantities with hat are eleven-dimensional ones.

After we have characterised the backgrounds by computing their central charges, the goal is to compare them with the central charges obtained from the putative dual field theories to these solutions, in the holographic limit. 
Some comments are in order. For instance, in the case of the field theory read off from the brane configuration in Table \ref{tableblabla}, we can achieve an anomaly free quiver field theory following the rules in \cite{Lozano:2019jza,Lozano:2019zvg}.
Nevertheless, this gives a central charge that is apparently changing due to the extra gauge and flavour group insertions. We would expect 
 cancelations among them that will give the same central charge as before the deformation, or that their contributions are sub-leading in the holographic limit. We will elaborate more on this in a forthcoming publication.
%We will elaborate more on this in a forthcoming publication.
%OTHER OPTION AFTER THE RULES OF []. Notice, the extra flavour and colour groups are apparently changing the central charge of the CFT. We however expect a cancelation that gives the same central charge, or that the contributions are sub-leading in the holographic limit. We will elaborate more on this in a forthcoming publication.
%A different perspective would be to think of the deformed solutions as obtained after placing the seed brane configuration on a ten dimensional flat space with CY$_2$ symmetry, the latter deformed by TsT. Following this line of reasoning, we would not actually be adding additional branes but deforming first the space that the seed brane configuration will later back-react.
%We will elaborate more on this in a forthcoming publication. 

   \section{Comments on supersymmetry and G-structure of the solutions}\label{secsupersy}
        In this section we will study the supersymmetries preserved by the supergravity solutions in eqs. (\ref{back2}), (\ref{back3}) and (\ref{defin11d}), (\ref{sol11ddeformed}), based on the explicit form of the Killing spinors of the original solution (\ref{back1}).
         The conventions we follow for supersymmetry are detailed in Appendix \ref{supersymmetry}. The solution in eq. (\ref{back1}) preserves small $\mathcal{N}=(0,4)$ supersymmetry by construction. In the conventional approach, this implies the existence
         of two algebraic conditions on the ten-dimensional Majorana-Weyl (MW) spinor ensuring the vanishing of the supersymmetry variations.
               
        In order to see this explicitly, we decompose the ten-dimensional gamma matrices as follows
\begin{equation}
\Gamma^{\alpha}=\sigma_1\otimes\rho^{\alpha}\otimes \mathbb{I}, \quad \Gamma^{\mu}=\sigma_2\otimes  \mathbb{I}\otimes\gamma^{\mu}, 
\end{equation}
where $\rho^{\alpha}$ and $\gamma^{\mu}$ are three and seven-dimensional gamma matrices respectively
and the $\sigma_{i}$ are the usual Pauli matrices. In this notation the chirality matrix is $\Gamma^{11}=-\sigma_3\otimes\mathbb{I}\otimes\mathbb{I}$. 
After plugging the solution in eq. (\ref{back1}) (in the natural frame) into eq. (\ref{susy4}) we find the MW Killing spinor takes the form
 \begin{equation}\label{spinor}
 \epsilon_{1}=\begin{pmatrix} 1\\ 0\end{pmatrix}\otimes \zeta\otimes \chi_{1}, \quad \epsilon_{2}=\begin{pmatrix} 0\\ 1\end{pmatrix}\otimes \zeta \otimes \chi _{2},
 \end{equation}
 where $\zeta$ is the AdS$_3$ Killing spinor and 
 \begin{equation}\label{spinor7d}
 \chi_{_{1,2}}=e^{\frac{A}{2}}e^{i\frac{ \theta}{2}\sigma_2\gamma^6}e^{\frac{\varphi}{2} \,\gamma^{56}} e^{-\frac{1}{2}\arctan 2\frac{\sqrt{h_4 h_8}}{u'}\gamma^{56}\sigma_3}\chi_{_{(0)1,2}}, \qquad A=\frac{1}{2}\log\frac{u}{\sqrt{h_4h_8}}, 
 \end{equation}
a seven-dimensional spinor satisfying the projection conditions 
\begin{equation}
\gamma^{78910}\chi_{_{1,2}}=-\chi_{_{1,2}},\qquad  \frac{1}{\sqrt{f_1}}\left(2\sqrt{h_4 h_8}\gamma^{4}\sigma_1+u'\gamma^{456}i\sigma_2\right)\chi_{_{1,2}}=\chi_{_{1,2}}
\end{equation}
where $4,5,6$ and $7\ldots 10$ are flat indices corresponding to the $\rho$, S$^2$
 and T$^4$ directions respectively. The purpose of this section, is to find 
the number of spinor components which are compatible with the TsT transformation. 

Since the deformation involves a sequence of T dualities, a condition for preserving Killing spinors reduces to their invariance by the action of the Kosmann-Lie derivative along the Killing vector $K$ 
associated to the isometric direction we picked to perform the duality $\mathcal{L}_{_{K}}\epsilon=0$, where $\epsilon$ is the Killing spinor of the un-dualised solution. By considering $K=\partial_{y}$, the above condition reduces to $\partial_{y}\epsilon=0$ \cite{Kelekci:2014ima}. 
Moreover, invariance under the change of coordinates in the second
 direction also requires independence of it on the spinor. Therefore, supersymmetry is compatible with TsT transformations as long as the spinor is uncharged under the directions used for the transformation \cite{Orlando:2018kms}.

For the first solution in eq. (\ref{back2}), there is a residual U(1)$_{\varphi}$ which we may think of as a candidate R-charge for $\mathcal{N}=(0,2)$ preserved supersymmetry. 
However, the spinors (\ref{spinor}) are charged under this coordinate and T duality along this direction will project out this dependence.
 The residual U(1)$_{\varphi}$ is therefore a global symmetry and supersymmetry is completely broken. In other words, compatibility with the TsT transformation imposes the projection condition $\gamma^{56}\chi_{_{1,2}}=0$ breaking all supersymmetries. Despite the breaking of supersymmetries, this solution is interesting in its own since it still solves the BPS condition in eq. (\ref{conds}).

For the second marginally deformed solution in eq. (\ref{back3}), the spinor is independent of the T$^4$ directions, so we ensure supersymmetry is fully preserved. To be more precise, working with the supersymmetry
transformations for the solution in eq. (\ref{back3}),  we find 
\begin{equation}
\begin{split}
\delta\tilde{\lambda}_{1}=&e^{-\arctan\lambda \sqrt{\frac{h_4}{h_8}}\gamma^{9 10}}\delta\lambda_{1},\qquad \delta\tilde{\lambda}_{2}=\delta\lambda_{2},\\ 
\delta\tilde{\psi}_{\mu \, 1}=&e^{-\arctan\lambda \sqrt{\frac{h_4}{h_8}}\gamma^{9 10}} \delta\psi_{\mu \,1}, \quad \delta\tilde{\psi}_{\mu \,2}=\delta\psi_{\mu \,2},
\end{split}
\end{equation}
where tilde denotes fields after the transformation, provided we identify 
\begin{equation}\label{spinortst}
\tilde{\chi}_{_1}=e^{-\arctan\lambda \sqrt{\frac{h_4}{h_8}}\gamma^{9 10}}\chi_{_1}, \qquad \tilde{\chi}_{_{2}}=\chi_{_2},
\end{equation}
ensuring supersymmetry is preserved as the original solution does. This is along the lines of the generic result in \cite{Orlando:2018kms}, which in addition showed that the entire information of the transformation is encoded in an antisymmetric 
bi-vector associated to classical r-matrices solving the Yang-Baxter equation. 

%for the pourpose of finding the identity structure /SU(3) one can start from scratch by taking the rotated 7d spinors and construct the bispinors directly. An easier way will be to directly identify the way the bispinors of the seed solutions transform. This is easily obtained from the way we obtain the RR fluxes of the solution starting with the seed ones. Namely: Then it would be possible to identify the way the geometric structures changed. For SU(3) for instance they will be different J and \Omega with a possible different angle rotation. We leave this explicit analysis for a forthcoming publication. For this, we need to parametrise the polyforms in terms of the structure quantities. 
Let us now turn to the G-structure characterising the above background. To begin with, the solutions  in \cite{Lozano:2019zvg} were constructed by imposing that 
 they support an SU(2) structure on the five-dimensional internal space M$_5$ transverse to AdS$_3\times$ S$^2$. 
 %Geometric conditions implied by supersymmetry constraint M$_5$ such that the four-dimensional subspace corresponds to either (conformal) CY$_2$ or K$_3$ manifolds defining. Recall in the present  work we have focused on the former case where the solutions in addition have CY$_2$ as a symmetry. 
 For these solutions, this implies that the internal five-dimensional spinors are globally parallel. The deformed Killing spinors break the above condition, each of which defining an SU(2) structure, the intersection of which gives an identity structure. To be more precise, given the rotation of the internal spinor under TsT eq. (\ref{spinortst}), the transformed MW spinor takes the form (\ref{spinor}) with the internal spinor transformed accordingly $\chi_{1,2} \rightarrow \tilde{\chi}_{1,2}$. 
In addition, the seven-dimensional spinor can be further decomposed into S$^2\times$ M$_5$ factors according to eq. (\ref{spinor7d}). Namely, $\chi_{_{1,2}}=e^{\frac{A}{2}}\xi\otimes \eta_{_{1,2}}$, where $\xi$ is a 
Killing spinor on the S$^2$ charged under SU(2)$_{R}$.  An SU(2) structure on M$_5$ implies\footnote{In a common basis the spinors can be written as $\eta_1=\eta$ and $\eta_2=a\eta+b\eta^{c}+\frac{c}{2}\overline{\omega}\eta$, where $\omega$ is a complex one-form. The case of SU(2) structure sets $c=0$. In addition, without loss of generality, we can choose for these solutions $b=0$. {\it class} I solutions are further characterised by $a$=1.}
 \begin{equation}
 \eta_1=\eta,\quad \eta_2=\eta. 
 \end{equation}
   Using (\ref{spinortst}) the TsT MW spinors are given by
   \begin{equation}
 \epsilon_{1}=\begin{pmatrix} 1\\ 0\end{pmatrix}\otimes \zeta \otimes \tilde{\chi}_{1}, \quad \epsilon_{2}=\begin{pmatrix} 0\\ 1\end{pmatrix}\otimes \zeta \otimes \tilde{\chi _{2}},
 \end{equation}
 where, using the 2+5 split of the internal spinor $\chi$, we find
 \begin{equation}
 \tilde{\eta}_1 =\frac{1}{\sqrt{1+\lambda^2\frac{h_4}{h_8}}}(\eta-\lambda\sqrt{\frac{h_4}{h_8}}(\gamma^{9 10}_{_{(5)}}\eta)),\qquad \tilde{\eta}_2 =\eta,
 \end{equation}
therefore the spinors $\tilde{\eta}_{1,2}$ are nowhere parallel defining a point-dependent SU(2)$\times$SU(2) structure, that we will refer to it as dynamical. 
 This is then described in terms of the largest common subgroup, which then defines a dynamical identity structure.
  Notice we could have also analysed the G-structure of the solution in terms of the seven-dimensional spinor. In this case the seed solution 
 supports an SU(3) structure. It would then be possible to understand the fate of the seven-dimensional G-structure following \cite{Dibitetto:2018ftj}. The analysis of this section suggest this may 
 give a dynamical SU(3) structure, and will provide a new example of AdS$_3$ solutions with dynamical SU(3) structure. We plan to report on this in a forthcoming publication.
 
 For the eleven dimensional solutions in eqs. (\ref{defin11d}) and (\ref{sol11ddeformed}) the preserved Majorana Killing spinors can be ascertained just as we did for the ten dimensional case. Namely, the preserved Killing spinors are those which are
independent of the directions along which we performed the transformation. Using the relation between the eleven and ten-dimensional spinors (\ref{spinorel}) together with (\ref{gravitinovar}) we easily see that small 
$\mathcal{N}=(0,4)$ supersymmetry is preserved.  Once again whenever the deformation parameter is turned offƒ
we recover the undeformed Majorana Killing spinor defining an SU(2) structure. In the case at hand we have a dynamical identity structure instead. 

 % Before closing this section, let us briefly comment on yet another way to see that supersymmetry is preserved. This is based on the way the TsT transformation acts on the bispinors  (or polyforms under the Clifford map)
 %used to construct the solutions. In this formalism,  the supersymmetry variations are written in terms of differential conditions on these polyforms constructed upon the internal Majorana spinors of the solution 
 %(for details see Section 2 of \cite{Lozano:2019emq}). In this vein, the goal is to ascertain the way these polyforms transform under TsT. Such transformation can be obtained from the way the RR sector of the deformed solution is extracted from the original ones (up to dilaton factors). 
 %We find 
  %$
 %\tilde{\slashed{\Psi}}_{\pm}=\slashed{\Psi}_{\pm}\Omega,
%$
 %where $\bold{\slashed{\Psi}}_{\pm}$ is the bispinor defining the SU(2)/SU(3) structure solutions. It would then be possible to prove that the transformed bispinors together with the fields defining the solution in eq. (\ref{back3}) do satisfy the
 %equations for $\mathcal{N}=1$ supersymmetry found in\footnote{Notice the method used to construct the seed solutions boils down to focusing on a $\mathcal{N}=1$ subsector of the solution. The rest can be generated with the action of the SU(2)$_R$ symmetry. This is respected by the solution we are discussing and therefore $\mathcal{N}=(4,0)$ supersymmetry is ensured.} \cite{Dibitetto:2018ftj}.  
 
 %In case we have the spinors written in terms of either the SU(2)/SU(3) quantities, the omega transformation will tell us how those quantities will transform

\section{Conclusions}\label{conclusions}
In this paper we have presented new solutions in massive IIA and eleven-dimensional supergravity obtained via TsT transformations and the analog in eleven dimensions. The solutions obtained 
preserve small $\mathcal{N}=(0,4)$ supersymmetry and support a dynamical identity structure on the five-dimensional internal submanifold of the solution, as long as we do not use the azimuthal angle inside the S$^2$
in the procedure. The new backgrounds in ten and eleven dimensional supergravity constitute a whole family of solutions parametrised by the deformation parameter $\lambda$ and linear functions satisfying the conditions in eq. (\ref{conds}). 
To the best of our knowledge, a complete classification of these solutions is still missing in the literature. One can in principle follow the same procedure as the one outlined in \cite{Lozano:2019emq} for the SU(2) structure case. That is to say construct bispinors out of seven-dimensional spinors supporting a (dynamical) identity structure in the internal five-dimensional space and obtain geometrical constraints in the form of the solution from the differential conditions implied by supersymmetry. Moreover, in terms of seven-dimensional G-structure, the seed solutions support an SU(3) structure. The analysis we followed in Section \ref{secsupersy} suggests this becomes a dynamical SU(3) structure after the transformation. 
Progress on classification of AdS$_3$ geometries supporting a dynamical SU(3) structure was recently reported in \cite{Passias:2020ubv}. 

For the ten-dimensional solutions, we studied the Page charges and associated brane configurations. We showed that depending on the two-torus chosen, the deformation adds either colour or flavour branes or both to the seed configuration.
Holographically, The backgrounds obtained correspond to marginal deformations of the SCFT dual to the seed solutions.  We verified this by computing the central charge of the deformed backgrounds, showing 
they are the same before and after the transformation. In the field theory side side, we can engineer a dual quiver quantum field theory with the information obtained from the Hannany-Witten brane set-ups associated to the solutions.
 The specification of the dual quantum field theories and more field theory aspects of the solutions are left for a forthcoming publication.  

%However, we ponted out that 
%via the quantised charges of the solutions. In particular,  for the solution in eq. (\ref{back3})
%we noticed the deformation adds extra colour and flavour D branes. If the brane-set up we read off from the information 
%quantised charges works for the deformed case then the extra fields change the range of the gauge and colour groups in the quiver. 
%Even thought we can achieve an anomaly free CFT the central charge apparently changes, in contradiction with the holographic computation. The specification of the CFT and more field theory aspects of the solutions are left for a forthcoming publication.

\acknowledgments

I am indebted to Carlos N\'u{\~n}ez for many useful discussions. I also thank Yolanda Lozano, Niall Macpherson, Anayeli Ramirez and Stefano Speziali for comments and correspondence.  

\appendix
\section{Supersymmetry conventions}\label{supersymmetry}

In this appendix we will set the conventions for supersymmetry. We find useful to review how to obtain the ten-dimensional supersymmetry variations 
from the eleven dimensional one by dimensional reduction.
 
Let us start with the supersymmetry variation for the gravitino in eleven-dimensional supergravity. It is
\begin{equation}\label{gravi}
\delta\hat{\psi}_{_{M}}=\left[\nabla_{M}+\frac{1}{288}\left(\hat{\Gamma}_{M}^{N_1\ldots N_4}-8\delta_{M}^{N_1}\hat{\Gamma}^{N_2 N_3 N_4}\right)\hat{F}_{N_1\ldots N_4}\right]\hat{\epsilon},
\end{equation}
where $\nabla_{M}=\partial_{M}+\frac{1}{4}\omega_{M}^{\,\,\,\, BC}\hat{\Gamma}_{BC}$.  We will study the reduction of the above supersymmetry variation along $x^{11}=z$.  From now on 
objects with hat will denote eleven dimensional quantities.
To proceed, we use the usual ansatz  for the string frame metric and three-form potential, which read 
 \begin{equation}
    \begin{split}\label{anpp}
    ds_{11}^2=&=\eta_{AB}E^{A}_{N_1}E^{B}_{N_2}dx^{N_1}dx^{N_2}=e^{-\frac{2}{3}\Phi}ds_{10}^{2}+e^{\frac{4}{3}\Phi}(dz+A_{1})^2,\\
    &\qquad \qquad C_3=A_3+B_2\wedge dz,
    \end{split}
    \end{equation}    
    where $\Phi$ is the dilaton and $M,N=(\mu,z)$, $A=(a,z)$ are curved and tangent space indices respectively.  The spin-connection components of the above geometry are thus given by 
    \begin{equation}
    \begin{split}
    \tilde{\omega}_{\,\,\, a b}= &e^{\frac{\Phi}{3}}(\omega_{\,\,\, a b}-\frac{2}{3}\eta_{c [a}\partial_{b]}\phi e^{c})-\frac{1}{2}e^{\frac{4}{3}\Phi}F_{a b}e^{z},\\
     \tilde{\omega}_{\,\,\, z a}=&\frac{2}{3}e^{\frac{\Phi}{3}}\partial_{a}\Phi e^{z}+\frac{1}{2}e^{\frac{4}{3} \Phi}F_{a b}e^{b},
    \end{split}
    \end{equation}
    where $F_2=2\partial A_1$. The dimensional reduction of the eleven-dimensional gravitino $\hat{\psi}_{_{M}}$ generates a ten-dimensional gravitino and the dilatino as follows
\begin{equation}\label{gravitinovar}
 \hat{\psi}_{\mu}=e^{\frac{\Phi}{6}}(\psi_\mu-\frac{1}{6}\hat{\Gamma}_{\mu}\lambda), \quad \hat{\psi}_{z}=\frac{1}{3}e^{\frac{\Phi}{6}}\hat{\Gamma}^{z}\lambda. 
\end{equation}

In the same vein, the $\hat{F}_4=4\partial C_3$ field strength contains two pieces the components of which are $\hat{F}_{\alpha\beta\gamma\delta}$ and $\hat{F}_{\alpha\beta\gamma z}$. Using flat indices we identify
\begin{equation}
\begin{split}
\hat{F}_{abcd}=&E_{a}^{N_1} E_{b}^{N_2} E_{c}^{N_3} E_{d}^{N_4}\hat{F}_{N_1 N_2 N_3 N_4}=4e^{\frac{4}{3}\phi}\left(\partial_{[a} A_{bcd]}-A_{[a} H_{bcd]}\right)=e^{\frac{4}{3}\Phi}F_{abcd}, \\
\hat{F}_{abc z}=&E_{a}^{N_1} E_{b}^{N_2} E_{c}^{N_3} E_{z}^{z}\hat{F}_{N_1 N_2 N_3 z}=e^{\frac{\Phi}{3}} H_{abc},
\end{split}
\end{equation} 
where $H=3\partial B_2$. Moreover, the dimensional reduction of  eq. (\ref{gravi}) generates the terms

\begin{align}
\delta \hat{\psi}_{z}=&\partial_{z}\hat{\epsilon}+\frac{1}{4}\left(\omega_{z\,\, a b}\hat{\Gamma}^{ab}+2\omega_{z\,\, z a}\hat{\Gamma}^{z a}\right)\hat{\epsilon}+\frac{1}{288}\left(\hat{\Gamma}_{z}^{A B C D}-8\delta_{z}^{A}\hat{\Gamma}^{B C D}\right)\hat{F}_{A B C D}\nonumber\\
=&\frac{1}{3}e^{\Phi/6}\hat{\Gamma}^{z}\left(\slashed{\partial}\phi-\frac{3}{4\cdot 2!}e^{\Phi}\slashed{F}_{2}\hat{\Gamma}^{z}-\frac{1}{12}\slashed{H}_{3}\hat{\Gamma}^{z}+\frac{1}{4\cdot  4!}e^{\Phi}\slashed {F}_{4} \right)\epsilon, \\
\delta\hat{\psi}_{a}=&E_{a}^{\mu}\partial_{\mu}\hat{\epsilon}+\frac{1}{4}\omega_{a BC}\hat{\Gamma}^{BC}\hat{\epsilon}+\frac{1}{288}\left(\hat{\Gamma}_{a}^{BCDE}-12\delta_{a}^{B}\hat{\Gamma}^{CDE}\right)\hat{F}_{BCDE}\hat{\epsilon}\nonumber\\
=&e^{\frac{\Phi}{6}}\left( (e_{a}^{\mu}\partial_{\mu}+\frac{1}{4}\omega_{a b c}\hat{\Gamma}^{b c}-\frac{1}{6}\partial_{a}\Phi -\frac{1}{6}\hat{\Gamma}_{a}^{c}\partial_{c}\Phi+\frac{1}{4}e^{\Phi} F_{a c}\hat{\Gamma}^{c z})  \right.\\
& \left. +\frac{1}{288}e^{\Phi}\hat{\Gamma}_{a}\slashed{F}_4-\frac{1}{24} e^{\Phi}\hat{\Gamma}^{bcd} F_{a bcd}+\frac{1}{72}\hat{\Gamma}_{a}\slashed{H}_3 \hat{\Gamma}^{z}-\frac{1}{8}\slashed{H}_{\mu}\hat{\Gamma}^{z}\right)\epsilon \nonumber,
\end{align}
where we have introduced the notation $\slashed{F}=F_{\mu_{1}\ldots\mu_n}\Gamma^{\mu_1\ldots\mu_n}$ and the relation between eleven and ten dimensional spinors 
\begin{equation}\label{spinorel}
\hat{\epsilon}=e^{-\frac{\Phi}{6}}\, \epsilon. 
\end{equation}

We then decompose the eleven-dimensional Gamma matrices in terms of the ten-dimensional ones in the following way
\begin{equation}\label{gammasplit2}
\hat{\Gamma}^{a}=\Gamma^{a}\sigma_{1}, \quad \hat{\Gamma}^{z}=\sigma_3, \quad a=1,\ldots 10,
\end{equation}

which is related to the decomposition of the ten-dimensional Majorana spinor into its chiral components
\begin{equation}
\epsilon=\begin{pmatrix}\epsilon_{1}\\
\epsilon_{2}\\
\end{pmatrix},
\end{equation}
satisfying $\Gamma^{11}\epsilon=-\sigma_3\epsilon$.  
Using (\ref{gravitinovar})-(\ref{gammasplit2}) we then identify the supersymmetry variations for the dilatino and gravitino. For non-zero Romans mass, the expressions obtained can be slightly generalised to
\begin{equation}
\begin{split}\label{susy4}
\delta\lambda=&\slashed{\partial}\Phi\epsilon-\frac{1}{2\cdot 3!}\slashed{H}\sigma_3\epsilon+\frac{e^{\Phi}}{4}\left(5 F_0\sigma_1+\frac{3}{2 !}\slashed{F}_2 i\sigma_2+\frac{1}{4!}\slashed{F}_4\sigma_1\right)\epsilon,\\
\delta\Psi_{\mu}=&\nabla_{\mu}\epsilon-\frac{1}{8}\slashed{H}_{\mu}\sigma_3\epsilon+\frac{e^{\Phi}}{8}\left(5 F_0\sigma_1+\frac{1}{2 !}\slashed{F}_2 i\sigma_2+\frac{1}{4!}\slashed{F}_4\sigma_1\right)\Gamma_{\mu}\epsilon.
\end{split}
\end{equation}

\end{document}